\documentclass[conference]{IEEEtran}
\IEEEoverridecommandlockouts
\usepackage[utf8]{inputenc}
\usepackage[style=ieee,minnames=3,maxnames=3]{biblatex}
\usepackage{amsmath,amssymb,amsfonts}
\usepackage{algorithmic}
\usepackage{graphicx}
\usepackage{textcomp}
\usepackage{xcolor}
\def\BibTeX{{\rm B\kern-.05em{\sc i\kern-.025em b}\kern-.08em
    T\kern-.1667em\lower.7ex\hbox{E}\kern-.125emX}}

\usepackage[nolist,nohyperlinks]{acronym}
\usepackage{booktabs}
\usepackage[disable]{easy-todo}
\usepackage{mathtools}
\usepackage{braket}
\acrodef{mu}[MU]{machine units}
\acrodef{dag}[DAG]{directed acyclic graph}
\acrodef{artiq}[ARTIQ]{advanced real-time infrastructure for quantum physics}
\acrodef{nisq}[NISQ]{noisy intermediate-scale quantum}
\acrodef{dsl}[DSL]{domain-specific language}
\acrodef{jit}[JIT]{just-in-time}
\acrodef{rb}[RB]{randomized benchmarking}
\acrodef{dds}[DDS]{direct digital synthesis}
\acrodef{dac}[DAC]{digital-to-analog converter}
\acrodef{adc}[ADC]{analog-to-digital converter}
\acrodef{awg}[AWG]{arbitrary waveform generator}
\acrodef{fpga}[FPGA]{field-programmable gate array}
\acrodef{yb171}[${}^{171}$Yb$^+$]{Ytterbium 171}
\acrodef{pmt}[PMT]{photomultiplier tube}
\acrodef{api}[API]{application programming interface}
\acrodef{rpc}[RPC]{remote procedure call}
\acrodef{mw}[MW]{microwave}
\acrodef{bb1}[BB1]{broadband}
\acrodef{sk1}[SK1]{Solovay-Kitaev}
\acrodef{spam}[SPAM]{state preparation and measurement}
\acrodef{cw}[CW]{continuous wave}
\acrodef{rtio}[RTIO]{real-time I/O}
\acrodef{sqst}[SQST]{single-qubit state tomography}
\acrodef{gst}[GST]{gate set tomography}
\acrodef{1d}[1D]{one-dimensional}
\acrodef{2d}[2D]{two-dimensional}

\acrodef{dax}[DAX]{Duke ARTIQ extensions}
\acrodef{staq}[STAQ]{software-tailored architecture for quantum co-design}
\acrodef{rc}[RC]{red chamber}
\begin{document}

\title{Modular Software for\\Real-Time Quantum Control Systems
% \thanks{Identify applicable funding agency here. If none, delete this.}
}

\author{
\IEEEauthorblockN{
Leon~Riesebos\IEEEauthorrefmark{1}\IEEEauthorrefmark{2},
Brad~Bondurant\IEEEauthorrefmark{1},
Jacob~Whitlow\IEEEauthorrefmark{1},
Junki~Kim\IEEEauthorrefmark{3},
Mark~Kuzyk\IEEEauthorrefmark{1},
Tianyi~Chen\IEEEauthorrefmark{4},
Samuel~Phiri\IEEEauthorrefmark{1},\\
Ye~Wang\IEEEauthorrefmark{1},
Chao~Fang\IEEEauthorrefmark{1},
Andrew~Van~Horn\IEEEauthorrefmark{1},
Jungsang~Kim\IEEEauthorrefmark{1} and
Kenneth~R.~Brown\IEEEauthorrefmark{1}
}
\IEEEauthorblockA{
\IEEEauthorrefmark{1}Department of Electrical and Computer Engineering,
Duke University, NC 27708, USA
}
\IEEEauthorblockA{
\IEEEauthorrefmark{2}Email: leon.riesebos@duke.edu
}
\IEEEauthorblockA{
\IEEEauthorrefmark{3}SKKU Advanced Institute of Nanotechnology (SAINT) and Department of Nanoengineering,\\
Sungkyunkwan University, Suwon 16419, Korea
}
\IEEEauthorblockA{
\IEEEauthorrefmark{4}Department of Physics,
Duke University, NC 27708, USA
}
}

\maketitle

\begin{abstract}
Real-time control software and hardware is essential for operating quantum computers.
In particular, the software plays a crucial role in bridging the gap between quantum programs and the quantum system.
Unfortunately, current control software is often optimized for a specific system at the cost of flexibility and portability.
We propose a systematic design strategy for modular real-time quantum control software and demonstrate that modular control software can reduce the execution time overhead of kernels by 63.3\% on average while not increasing the binary size.
% When allowing buffering, the modular control software can reduce execution time overhead by 88.7\% on average.
Our analysis shows that modular control software for two distinctly different systems can share between 49.8\% and 91.0\% of covered code statements.
To demonstrate the modularity and portability of our software architecture, we run a portable randomized benchmarking experiment on two different ion-trap quantum systems.

% 100 word limit (not for QCE22?)

\end{abstract}


% Content
\acresetall
\section{Introduction}
\label{sec:introduction}

\begin{figure}
    \centering
    \includegraphics[width=\linewidth]{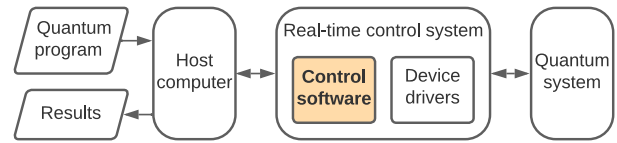}
    \caption{A real-time control system bridging the gap between the quantum program and the quantum system.}
    \label{fig:overview}
\end{figure}

The field of quantum computing is rapidly evolving in the areas of software and hardware. On the software side, quantum programming languages and compilers are becoming more available and feature-rich \cite{svore2018q, cross2021openqasm, Qiskit, fu2020quingo, Steiger2018projectqopensource, cirq_developers_2021_4586899}. At the same time, quantum hardware is becoming increasingly powerful with recent systems demonstrating computations on tens of qubits \cite{arute2019quantum, ryan2021realization, postler2021demonstration, wang201816, pogorelov2021compact, acharya2022suppressing, 810291}.
An often underexposed area in the field of quantum computing is the control software and hardware that bridges the gap between the quantum program and the targeted quantum system.
Recent papers \cite{arute2019quantum, kim2020hardware, blok2020quantum, pogorelov2021compact} have shown that current state-of-the-art quantum systems already require tens to hundreds of devices to be controlled with high precision and strict real-time requirements, proving to be a significant challenge for the control system.
Existing control hardware as described in \cite{bourdeauducq_2016_51303, negnevitsky2018feedback, ioncontrol, fu2019eqasm, ryan2017hardware} provides the required real-time control of devices, but it is up to the real-time control software to close the remaining gap between device-level control hardware and quantum programs as illustrated in Figure~\ref{fig:overview}.

Control software for quantum systems that runs on the real-time controller is similar to high-performance and resource-constrained embedded software. The software is responsible for real-time control of a set of devices while capturing and processing data simultaneously.
Real-time controlled devices include \ac{dds} devices, digital I/O, and \acp{dac}.
Additionally, the real-time control system for a quantum system functions as a coprocessor and maintains a connection with a host computer.
Due to the performance requirements and the strong dependence between the software and hardware, real-time control software is often tailored for a specific quantum system at the cost of flexibility and portability.
Since quantum computing is still an emerging technology, most quantum systems are unique. As a result, real-time control software is often redeveloped for each system which causes significant development overhead.

In this paper, we propose a systematic design strategy for real-time quantum control software.
We present an open-source software framework for the \ac{artiq} open-source software and hardware ecosystem \cite{bourdeauducq_2016_51303, Kasprowicz:20} to apply our design concepts to real-time quantum control software.
Our framework supports the development of modular control software to enhance flexibility and portability on the level of real-time system code. Portability on the application level is achieved by introducing software abstractions. We show that modular control software developed with our framework can reduce the execution time overhead of real-time software and achieve high degrees of code portability.
Reduced execution time overhead is achieved by fine-grained timing management and data offloading features of the modular software.
We demonstrate the capabilities of our software framework by running a portable \acl{rb} experiment on two different ion-trap quantum systems that are fully controlled by software based on our framework.

The remainder of this paper is structured as follows. Related work is discussed in Section~\ref{sec:related}. In Section~\ref{sec:architecture} we present our design strategy and modular software architecture for real-time quantum control software.
Our performance analysis and code portability analysis can be found in Section~\ref{sec:evaluation} and~\ref{sec:portability}, respectively.
In Section~\ref{sec:experiments} we present experimental results from two ion-trap quantum systems. We conclude our paper in Section~\ref{sec:conclusion}.

\section{Related work}
\label{sec:related}

% https://github.com/sinara-hw/meta/wiki/competition

Control software for \ac{artiq} systems can be developed without the use of our proposed framework. Programmers will have access to a classically complete programming environment, basic data storage utilities, and device drivers to program real-time devices. However, \ac{artiq} is set up as a fully generic control system and no utilities are provided to support modular software development. As a result, control software is often tightly coupled to the hardware and needs to be completely redeveloped for each system. Especially the real-time timing of the software is often highly dependent on the controlled devices. The tight coupling of the hardware and software makes it difficult to change devices in existing systems since any modification likely introduces timing issues.
At this moment, we are not aware of any other frameworks or libraries to support the development of modular control software for \ac{artiq}.
Other real-time control systems similar to \ac{artiq}, such as M-ACTION \cite{negnevitsky2018feedback, pogorelov2021compact} and IonControl \cite{ioncontrol}, suffer from the same limitations.

QCoDeS \cite{qcodes} is a modular data acquisition framework mainly intended to orchestrate the setup and data collection of instruments and devices part of a quantum control system. Some of these instruments can have real-time features, but QCoDeS does not control instruments in real-time while an experiment runs. Instead, all code involving QCoDeS runs on the host computer in a Python environment. Real-time instruments are also much more coarse compared to \ac{artiq}. A single \ac{artiq} real-time controller with many real-time I/O devices would correspond to a single QCoDeS instrument.
The concepts behind QCoDeS show some similarities with the non-real-time components of the \ac{artiq} host environment. What sets \ac{artiq} apart from QCoDeS is its seamless integration and combination of real-time software within the host environment. \ac{artiq} puts the real-time controller in the center based on the principles of the accelerator model, while QCoDeS behaves more as a hypervisor for devices.
The concepts presented in this paper apply to low-level real-time control software and its interaction with the host, and QCoDeS is not involved in the former.
QCoDeS does have features for software modularity but these are limited to the instrument level without introducing any form of hierarchy.
Other software based on or derived from QCoDeS, such as PycQED \cite{pycqed}, is built on the same principles and has the same limitations.

Qiskit \cite{Qiskit} is an open-source library for creating, compiling, and executing quantum programs at the gate level. While it does allow users to execute quantum programs, Qiskit itself does not transparently connect to any device-level drivers and is not directly involved in the real-time control of devices when the compiled quantum program runs. Hence, Qiskit merely describes a circuit and is not directly part of the real-time control software that runs the circuit on hardware. Even the pulse-level control provided in Qiskit is an opaque abstraction over the device-level drivers. The same holds for related and similar tools such as OpenQASM \cite{cross2021openqasm}, OpenPulse \cite{mckay2018qiskit}, Q\#\cite{svore2018q}, and Cirq \cite{cirq_developers_2021_4586899}.

\acresetall
\section{Software architecture}
\label{sec:architecture}

\begin{figure}
    \centering
    \includegraphics[scale=0.8]{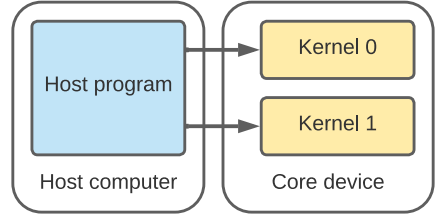}
    \caption{Schematic overview of the accelerator model with a host program and one or more kernels.}
    \label{fig:host_kernel}
\end{figure}

Our software architecture targets the \ac{artiq} open-source software and hardware ecosystem \cite{bourdeauducq_2016_51303, Kasprowicz:20} which is used by dozens of research groups and has deployed over 200 real-time control systems worldwide.
\Ac{artiq} follows the principles of the \emph{accelerator model} \cite{riesebos2019quantum, fu2019eqasm, svore2018q, fu2020quingo, nguyen2020extending, smith2016practical, chong2017programming, stone2010opencl}
where a program consists of a host program and one or more \emph{kernels}. The host program executes on a host machine and can offload the execution of kernels to an accelerator.
For \ac{artiq}, the host program runs on a classical computer while kernels run on the real-time control hardware as illustrated in Figure~\ref{fig:host_kernel}.
\Ac{artiq} kernels are classically complete and have access to real-time devices that interact with the quantum system (e.g. \ac{dds} devices, digital I/O, and \acp{dac}).
The real-time control hardware, referred to as the \emph{core device}, contains a classical CPU and a \ac{rtio} subsystem that schedules events for real-time devices on a timeline using a timeline cursor as described in \cite{fu2017experimental, fu2019eqasm}.
The \ac{artiq} system is programmed in a Python host environment and kernels are functions written in the \ac{artiq} \ac{dsl}, which is a Python-like language containing additional constructs for manipulating the timeline cursor and inserting events. \Ac{artiq} allows host and kernel code to be combined in a single file, and kernels are functions or methods decorated with the \verb;@kernel; decorator. When the host calls a kernel function, the host will invoke the \ac{artiq} compiler that compiles the kernel to a binary. The resulting binary is uploaded to and executed by the core device.
While the core device executes the kernel, the host serves any synchronous or asynchronous \acp{rpc} from the core device. \acp{rpc} are often used to stream real-time data to the host, access peripheral (i.e. non-real-time) devices, and offload computationally heavy tasks to the host.
Once the kernel finishes execution, the host program resumes execution.
The \ac{artiq} programming environment is set up generically and does not provide additional utilities for organizing real-time control software.
% In this section, we will describe how we apply object-oriented concepts to real-time control software written for an \ac{artiq} control system.
In this section, we will present our design principles for real-time control software written for an \ac{artiq} control system.

\begin{figure}
    \centering
    \includegraphics[width=0.9\linewidth]{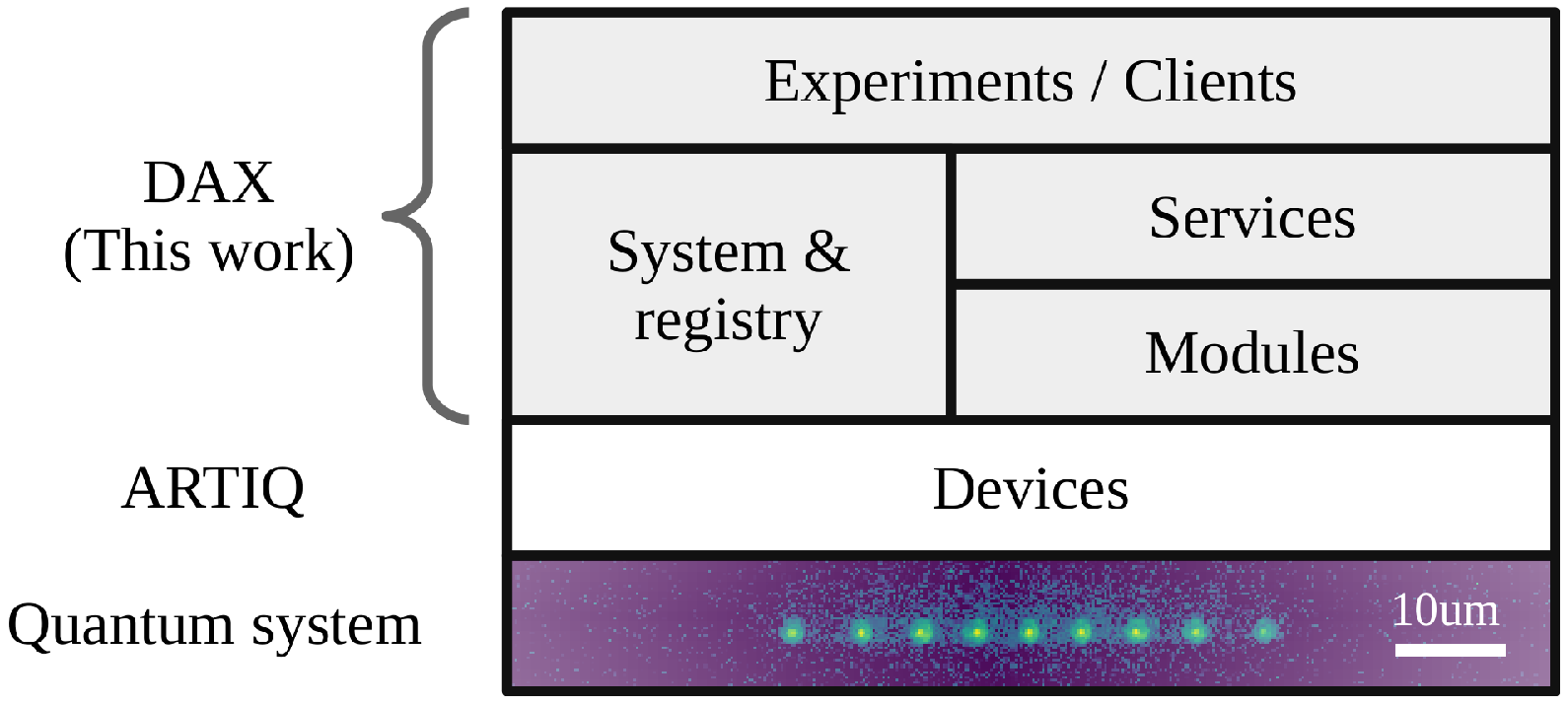}
    \caption{Schematic overview of the software components in our modular architecture controlling a quantum system, in this case, trapped atomic ions.}
    \label{fig:dax_arch}
\end{figure}

The first step towards code organization is to separate common functionality of the system (i.e. \emph{system code}) from experiment-specific routines (i.e. \emph{experiment code}). System code can be collected in a base class, and each experiment can inherit from such a class and add experiment-specific routines. This approach is already common practice for most \ac{artiq} experiments.
Our software architecture focuses on the development of modular system code. We propose to break the system code into two components: device organization with \emph{modules} and extensible system-wide functionality using \emph{services}. Additionally, we will introduce the notion of a central and searchable \emph{registry} in which all modules and services of a system are registered. To improve code portability, we introduce abstractions with \emph{interfaces} and \emph{clients}.
Figure~\ref{fig:dax_arch} shows an architectural overview of the different components, which we will describe in the remainder of this section.

\subsection{Modules, services, and the registry}
\label{sub:modules_services}

While most experiments require a large set of real-time devices to collaborate closely, subsets of devices that perform basic procedures often have a tighter relation from a control perspective. A subset of devices might have strict control or safety requirements independent from other devices in the system.
For example, two devices might always need to be switched simultaneously to achieve some desired functionality.
We introduce the concept of modules to group such a logical collection of devices.

A module is self-contained and controls zero or more devices that depend on each other to perform basic procedures.
For example, a detection module can contain devices required to apply a readout signal to the system together with the input devices that read the state of the qubits during detection.
To guarantee that modules are \emph{independent} from a control perspective, a device can only be assigned to a single module.
Each module has access to its own persistent data storage to store configuration and calibration data related to its operation.
The collective behavior of devices in a module can be described using module functions.
For example, a detection module could have a function that controls the readout signal and the input devices in parallel to perform a detection procedure.
Module functions are not solely used for collective device behavior and can also be used to manipulate devices separately, read or write configuration data, or update calibration parameters.

A system contains one or more modules that are organized in a tree structure. Every module in the system can contain zero or more sub-modules, and the root module is known as the \emph{system module}. Parent modules can access features of all their child modules, allowing hierarchical and transparent structuring of devices and functionality.
Because each device can only be assigned to a single module, modules in non-overlapping sub-trees of the system hierarchy are \emph{device-independent}. Two independent modules can be controlled in parallel, which means they can both add events to the \ac{rtio} event timeline without device conflicts.
Hence, the width of the module tree represents the amount of control- and device independence between different parts of the system. Device dependencies are encoded in the tree structure, and more independent modules lead to more available control parallelism in the system.
All modules are added to the central registry of the system such they can be easily found later.
Modules form the first level of system organization and introduce fundamental abstractions for device control and dependencies.

Modules introduce a straightforward device and system organization, but each separate module has limited power due to its local scope. Only the system module (i.e. the root module) can control all devices in the system, a requirement for most meaningful operations on the quantum system. With only modules, all system-wide functionality would have to be implemented in the system module, reducing the modularity of the software architecture. To overcome this issue, we introduce services as a technique to organize system-wide functionality.

A service is a component that can control multiple modules or even the whole system if desired. Through the registry, a service can obtain any modules required for its functionality. A single module in the system can be accessed by any number of services. The functions of services usually describe the collective behavior of multiple modules, and functionality can vary from short operations to lengthy procedures. If modules are device-independent, a service is allowed to control them in parallel.
% For example, an \verb;active_reset(); function of a service could call a detection module to perform a state detection and conditionally call other modules to actively reset qubits that are not in the $\ket{0}$ state.
Just as modules, services have access to their own persistent data storage, and every service is added to the central registry of the system.

Services are not limited to calling just modules. Through the registry, services can also find other services to use their functionality. Building services on top of other services allows transparent layering of increasingly complex system behavior. Hence, services are organized in a \ac{dag}.
Services contain powerful functions and enable the organization of system-wide functionality, but because of their system-wide control, services can not operate on the system in parallel with any other module or service as this could potentially lead to device-control conflicts.

\subsection{Interfaces and clients}
\label{sub:interfaces_clients}

System code can be organized with the concepts described in Section~\ref{sub:modules_services}, but because most quantum systems are unique, the majority of modules and services are still developed and optimized for a specific system.
To abstract system code, we introduce standardized interfaces.
Interfaces describe a set of functions that must be implemented by a module or service. A single module or service can implement multiple interfaces and a single interface can be implemented by multiple modules or services in a system.
For example, a gate interface could expose a set of functions that implement operations to perform quantum gates. Multiple instances of a gate interface within a single system could represent different gate implementations.
To prevent device- and control conflicts, interfaces should be considered as implemented by a service. Hence, an interface can not operate on a system parallel with any other module, service, or interface.

System code that implements one or more standardized interfaces can run portable experiments, which we call clients. Clients exclusively control a system through interfaces. A client is instantiated against a system, and the necessary interfaces will be obtained at runtime using the system registry. With clients, we can develop portable experiments that can run on different systems or different implementations of an interface in a single system.
The system code functions as middleware between the generic experiment described in the client and the system-specific implementation of the utilized interfaces.

\subsection{Implementation}
\label{sub:implementation}

We have implemented a software framework to support the development of real-time control software based on the presented concepts. The framework is part of our open-source library \ac{dax}
\cite{riesebos2021dax}  % Double blind
, which integrates tightly with the \ac{artiq} open-source software and hardware ecosystem.
\Ac{dax} implements a set of base classes that developers must inherit when defining modules and services. All these base classes provide direct access to data storage functions and the central registry of the system. 
When modules or services are instantiated, a unique hierarchical key and data storage location is assigned to the object based on the module tree or service \ac{dag}.
In addition, the \ac{dax} library contains standard modules and services with portable functionality that can be used by any system.
Such modules and services include functionality for processing measurement data, device-safety control, and common device control.
Additionally, we have developed a generic class that defines the standard control flow of a single- or multi-dimensional scanning-type calibration experiment.
Our \ac{dax} framework relies heavily on multiple inheritance to combine features of multiple classes, and fortunately, the Python host environment supports this well.

\Ac{dax} defines various interfaces that can be implemented by modules and services. The two interfaces of interest for this paper are the \emph{operation interface} and the \emph{data-context interface}. The operation interface contains functions for common gate-level quantum operations, including single- and two-qubit Clifford operations, arbitrary rotation gates, and qubit state preparation/measurement. The data-context interface is used to store and process obtained measurement results.
We developed clients to perform \ac{rb} \cite{magesan2011scalable, PhysRevLett.123.030503, epstein2014investigating} and \ac{gst} \cite{blume2013robust} which use the operation interface and the data-context interface to execute benchmark circuits. The \ac{rb} and \ac{gst} clients work with every system that uses \ac{dax}-based control software and implements the required interfaces.
Both clients are based on the open-source pyGSTi library \cite{pygsti} which is used to generate benchmarking circuits and analyze results.
Finally, we have defined an \ac{api} that can be used to write portable quantum programs in an \ac{artiq} environment given an operation interface and a data-context interface.
Using this \ac{api}, we implemented a program to perform \ac{sqst} \cite{schmied2016quantum}. Such portable programs can be executed by dynamically linking the program to the interfaces of a system using the program loader client we developed.

\section{Performance evaluation}
\label{sec:evaluation}

To evaluate the benefits and overhead of modular control software developed with the \ac{dax} framework, we re-implemented the control software for the \ac{staq} system, an experimental trapped-ion quantum processor~\cite{kim2020hardware}.
The real-time control hardware of \ac{staq} is based on a Kasli~2.0 controller \cite{Kasprowicz:20}, which is part of the \ac{artiq} hardware ecosystem.
The old \ac{artiq} control software for \ac{staq} is designed with a system-specific and monolithic architecture while the new modular control software is developed using our \ac{dax} framework.
In this section, we will compare the old control software with the new \ac{dax}-based control software. 

The old \ac{staq} control software separates system code from experiment code, but the system code has a monolithic architecture and is highly hardware dependent.
For example, code related to \ac{rtio} timing is highly dependent on the latencies introduced by the programming of real-time devices. Any changes to the devices will likely cause \ac{rtio} timing constraints violations throughout the code.
Hence, the old control software is not modular or portable.
We did make modifications to the old control software to optimize its performance and make the comparison to the new control software fair. All delays inserted on the event timeline used to compensate for device programming times larger than 200~us were reduced to 200~us or replaced by other efficient solutions that satisfy timing constraints. Such delays are often necessary to not violate any timing constraints of the \ac{rtio} system. The new control software also uses 200 us delays to compensate for device programming time, which has been found empirically to be sufficient. Minor bugs found in the old code were also fixed to ensure the old and new experiments are functionally equivalent.

\begin{figure}
    \centering
    \includegraphics[scale=0.5]{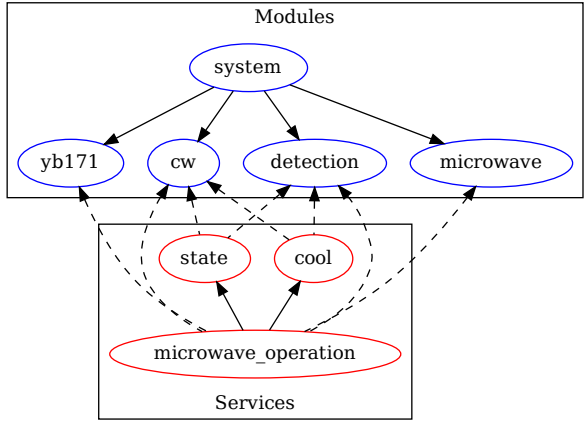}
    \caption{Subset of the \acs{staq} modules and services relevant for the \acl{mw} operation service.}
    \label{fig:staq_system}
\end{figure}

The new \ac{dax}-based system code for \ac{staq} is modular and organized in 11 modules and 11 services.
Most modules and services are system-specific, but two services use portable \ac{dax} data-processing modules, and one module extends a \ac{dax} module with safety-related functionality.
The \ac{dax}-based system code implements various \ac{dax} interfaces, including the data-context interface and four implementations of the operation interface. The new system code packs more features and complexity, including control over more real-time devices (increased from 23 to 35) and external devices (increased from 4 to 8).
Figure~\ref{fig:staq_system} shows a subset of the \ac{staq} modules and services relevant for the \acl{mw} operation service, which implements the \ac{dax} operation interface. Solid arrows show the tree structure and \ac{dag} dependencies for modules and services, respectively. The dashed arrows indicate modules that are directly used by services.
The \acl{mw} module controls a \ac{dds} to apply \acl{mw} pulses to the ions. The \acp{pmt} and lasers used for detection are controlled by the detection module. The \ac{cw} module controls various other lasers for cooling and pumping while the Yb171 module stores ion calibration data.
The cool service contains various subroutines for cooling ions while the state service implements the data-context interface and is used for state initialization and detection. Finally, the \acl{mw} operation service uses all the mentioned modules and services to perform \acl{mw} gates, qubit state preparation, and measurements.

We chose five relevant experiments with a single real-time kernel available in both the new and the old \ac{staq} control software for comparison.
The selected experiments include two \acf{mw} experiments (\acs{mw} freq/time), a qubit initialization experiment (qubit init), a tickle experiment (tickle), and an Ytterbium spectroscopy experiment (Yb spec).
All experiments are \ac{1d} scanning-type experiments and scan over 20 data points. The new control software utilizes the generic \ac{dax} scanning infrastructure while the old control software has defined scanning control-flow procedures as part of the system code.
Each experiment takes 100 samples per point except for Yb spec, which takes 30 samples per point. We ran each experiment with the same configuration using the old and new control software.
Additionally, we run each experiment using the new control software with buffering enabled. Buffering allows the real-time control software to schedule the operations for the next samples while the incoming data of earlier samples are kept temporally in hardware buffers. \Ac{artiq} supports such hardware buffers, but the real-time software must be designed appropriately to utilize them. Buffering can further increase the throughput and performance of kernels by reducing stalling time at the cost of increased latency between receiving and processing input events. None of the mentioned experiments are sensitive to the increased latency and will benefit from increased throughput. We configure a buffer size of 16 samples, which should be large enough to get the maximum performance gain achievable with buffering. The old control software does not include features for buffering.
We measured the execution time of the kernel with nanosecond precision using the real-time clock available in the Kasli controller (i.e. the core device).
An execution time measurement starts when the kernel starts execution, after the kernel binary is compiled and uploaded to the core device, and stops when the kernel finishes execution. Any \acp{rpc} from the core device to the host are included in the execution time measurement.
We will use the execution time measurements to calculate the overhead of the real-time software.
The kernel binary size is measured on the host at the output of the \ac{artiq} compiler and is used to calculate any binary-size overhead caused by our software framework.
All our measurements are performed with \ac{artiq} version~6.7659.c6a7b8a8 and the results are presented in Figure~\ref{fig:runtime_overhead} and~\ref{fig:kernel_size}.

\subsection{Execution time overhead}
\label{sub:overhead}

\begin{figure}
    \centering
    \includegraphics[width=\linewidth]{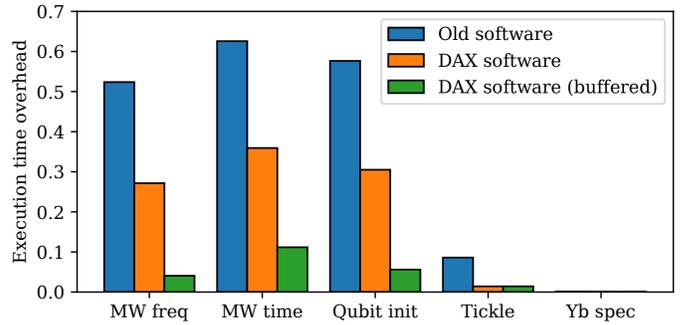}
    \caption{Kernel execution time overhead for the old control software and new \acs{dax}-based control software of the \acs{staq} system.}
    \label{fig:runtime_overhead}
\end{figure}
% Execution time overhead Old software [0.524 0.626 0.576 0.086 0.001]
% Execution time overhead AMX software [0.271 0.359 0.305 0.014 0.   ]
% Execution time overhead AMX software (buffered) [0.041 0.112 0.056 0.014 0.   ]

The results in Figure~\ref{fig:runtime_overhead} show the execution time overhead of the kernel for each experiment using the old and new \ac{dax}-based control software.
For each experiment, we calculate the minimal execution time~$t_\textit{min}$ based on the pulse lengths, detection times, and intentional wait times of the experiment. Given the measured execution time of an experiment~$t_\textit{exe}$, the execution time overhead is defined as $(t_\textit{exe} - t_\textit{min}) / t_\textit{min}$.
Figure~\ref{fig:runtime_overhead} shows that the old control software has an execution time overhead between~52.4\% and~62.6\% for the two \ac{mw} and the qubit init experiments. These experiments consist of relatively short and quick operations which increase the operation density and induce more strain on the \ac{rtio} subsystem.
% For reference, the lower bound execution times for the \ac{mw} and qubit init experiments are between 1400.23 and 1520.00 ms.
Any inter-sample execution overhead introduced by the real-time control software will quickly increase the total execution overhead.
The tickle experiment consists of slower operations resulting in a measured overhead of 8.6\% for the old control software. Any software overhead will be less significant due to the longer total duration of the experiment.
The Yb spec experiment has very slow operations and includes a 500 ms wait time for each sample. Any overhead introduced by the real-time control software will be negligible on the timescale of the experiment.

If we look at the results of the \ac{dax}-based control software (without buffering) in Figure~\ref{fig:runtime_overhead}, we see that the new control software significantly reduces the execution time overhead compared to the old control software. On average, the new control software reduces the execution time overhead by 63.3\% compared to the old control software. This average overhead reduction also includes the Yb spec experiment which already has a negligible execution time overhead for the old control software. When not including the Yb spec experiment, the average execution time overhead reduction is 55.4\%.
When we include buffering, we see that experiments with short and quick operations benefit the most (i.e. the two \ac{mw} and the qubit init experiments). Buffering reduces inter-sample overhead by scheduling multiple samples ahead before retrieving results. Hence, the experiments most affected by inter-sample overhead benefit the most from buffering. The execution time overhead of the tickle experiment is not further reduced by buffering. The new control software already reduced its overhead to 1.4\%, and inter-sample overhead does not appear to be a significant part of that. Compared to the old control software, the \ac{dax}-based control software with buffering enabled reduces execution time overhead by 88.7\% and 87.1\% on average with and without the Yb spec experiment, respectively.

We further analyzed our measurements to understand why the \ac{dax}-based control software performs better than the old control software. We attribute the reduced overhead to two main sources: timing management and data offloading. 
As mentioned earlier, real-time control software often inserts some delays on the event timeline to compensate for device programming times. The new control software groups devices in modules which in turn provides functions to manipulate those devices. The inserted delays can be optimized for each function which reduces the overhead. The old control software is less structured which often leads to larger worst-case delays or redundant delays to be inserted. Modular and well-designed real-time software allows us to insert more fine-grained delays, which reduces the total execution time overhead.
Modular software design also leads to code that is more flexible and robust to changes. When a module has any modifications to its real-time devices or their behavior, its function might need to be optimized again, but other modules and services are not affected by the change. If devices in a module completely change, a module might need to be redeveloped. Fortunately, if the function signatures of the new module are compatible with the old one, the modules could be swapped without affecting other parts of the system.
% For the \ac{staq} system, we have successfully modified and replaced modules and services in the past.

The second major contributor to overhead reduction is data offloading. Measurement data for an experiment is often offloaded to the host using asynchronous \acp{rpc} while the kernel is running. Such offloading can be very efficient and transfers parts of computational tasks from the kernel to the host while also reducing memory usage on the core device. The new control software uses portable \ac{dax} data-processing modules which are highly optimized to maximize the benefit of the data offloading. As a result, the complexity and execution time overhead of the kernel is reduced.

We would like to mention that better timing management and data offloading is also achievable with monolithic control software, but modular software makes it much easier. Devices will always be addressed through the functions of the module it is part of, making it easy to optimize the inserted delays for each scenario and improve timing management. For data offloading, the \ac{dax} data-processing module is portable between systems (see Section~\ref{sec:portability}) and only has to be developed and optimized once thanks to the modular software architecture.

\subsection{Kernel binary size}
\label{sub:kernel_size}

\begin{figure}
    \centering
    \includegraphics[width=\linewidth]{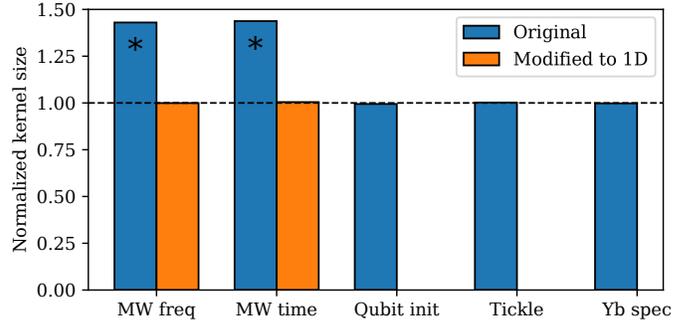}
    \caption{Kernel binary size of the new control software normalized to the kernel binary size of the old control software.
    \\{\Large $\ast$} In the new control software, the two \acf{mw} scan experiments merged into a single 2D scan experiment causing an increased kernel size.}
    \label{fig:kernel_size}
\end{figure}
% normalized_kernel_sizes [1.43  1.438 0.995 1.002 0.998] 1.172
% normalized_kernel_sizes_1d [0.999 1.005 0.995 1.002 0.998] 1.0

The results in Figure~\ref{fig:kernel_size} show the kernel binary size of the new control software normalized to the kernel size of the old control software. We see that for the two \ac{mw} experiments the kernel size is increased by 43.0\% to 43.8\% while for the other experiments the kernel size changed less than 1\%.
While all experiments are \ac{1d} scanning-type experiments, the two \ac{mw} experiments merged into a single \ac{2d} scanning experiment in the new control software. For our tests, we configured one dimension to be static to reduce the experiment to a \ac{1d} scan. While this will result in a functional \ac{1d} scan, the \ac{dax} scanning infrastructure still stores data for the static dimension for each point in the scan. Hence, the kernel binary size increases. We manually modified the new \ac{mw} experiment to a \ac{1d} scan for frequency and time storing the fixed value of the other dimension as a constant. When we measure the kernel binary size again, the difference with the old control software is less than 1\%.
From our results, we can conclude that the \ac{artiq} compiler works well with modular control software, and modular real-time software does not cause extra overhead that increases the kernel binary size.

\section{Code portability}
\label{sec:portability}

A principal benefit of modular control software is the potential for code portability between different quantum systems. The \ac{artiq} ecosystem already successfully abstracts real-time hardware with drivers and gateware to hide differences between hardware configurations or even hardware platforms.
The \ac{dax} framework tries to achieve portability and abstraction on a higher level, more specifically the system-level and application-level software.
On the system-level, generic \ac{dax} modules, services, and scanning infrastructure (see Section~\ref{sub:implementation}) allow portability of real-time control code between systems. Portability for application-level software is achieved by using interfaces and clients.

To evaluate the amount of code portability between two different systems, we have implemented \ac{dax}-based control software for a second experimental trapped-ion quantum system known as the \ac{rc} system~\cite{wang2020high}.
The real-time control hardware of \ac{rc} is based on a KC705 \cite{kc705} evaluation board with custom breakout boards that contain digital I/O and \ac{dds} devices.
The control software for the \ac{rc} system consists of 20 modules and 7 services. Two services use portable \ac{dax} data-processing modules and one module extends the \ac{dax} module for safety-related functionality. The \ac{rc} system code implements multiple \ac{dax} interfaces, including the data-context interface and two implementations of the operation interface. The system code controls 30 real-time devices and 1 external device. Notable is that the \ac{rc} system has more modules than the \ac{staq} system even though there are fewer real-time devices. The software of the \ac{rc} system was developed after that of the \ac{staq} system, and we learned it was better for modularity and portability to have a deeper system tree with more and smaller modules.

\begin{figure}
    \centering
    \includegraphics[scale=0.5]{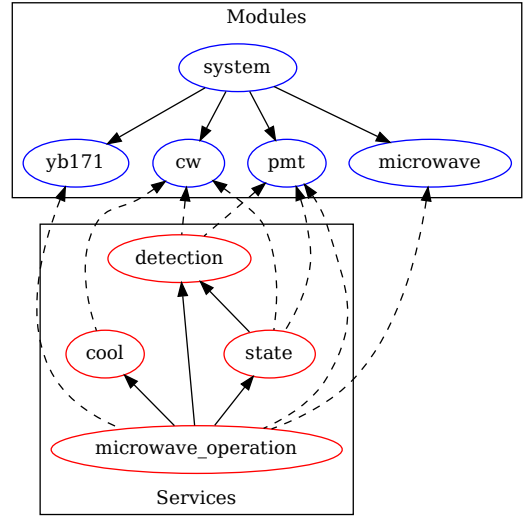}
    \caption{Subset of the \acs{rc} modules and services relevant for the \acl{mw} operation service.}
    \label{fig:red_chamber_system}
\end{figure}

Figure~\ref{fig:red_chamber_system} shows a subset of the \ac{rc} modules and services relevant for the \acl{mw} operation service, which implements the \ac{dax} operation interface.
% Not shown are the many submodules, for example, seven of them for the \ac{cw} module.
The graph looks very similar to the one for the \ac{staq} system shown in Figure~\ref{fig:staq_system} despite the real-time devices and controlled hardware being significantly different. The Yb171 and \acl{mw} modules are similar to their \ac{staq} counterparts while the main differences are found in the \ac{cw} and \ac{pmt} modules.
One key difference between the systems is that the detection laser shares an upstream master switch with other \acl{cw} lasers. Due to the master switch, it is impossible to control the detection laser independently from other lasers in the \ac{cw} module without potential conflicts. Hence, the detection laser is controlled by the \ac{cw} module and the \acp{pmt} are contained in an independent module. A detection subroutine now requires the \ac{cw} and \ac{pmt} module to work in parallel which is captured in the detection service. The remaining services in the \ac{rc} system are similar to their \ac{staq} equivalents.
Figure~\ref{fig:staq_system} and~\ref{fig:red_chamber_system} show that two systems with significantly different real-time control systems and devices can still have real-time control software with similar architectures. Modules and services can successfully abstract such differences.

To evaluate code portability between the \ac{staq} and \ac{rc} system, we disabled modules and services not relevant for \acl{mw} operations. We then run a set of six experiments on each system. Three are \ac{mw} calibration experiments and include a \ac{mw} frequency calibration, a \ac{mw} Ramsey frequency calibration, and a \ac{mw} gate experiment that executes a sequence of $X$ rotations to fine-tune the \acl{mw} Rabi gate time. These calibration experiments are hardware-specific and therefore have system-specific implementations. Two other experiments are the \ac{dax} clients for \ac{rb} and \ac{gst} that use the operation interface and the data-context interface. The last experiment is the portable \ac{sqst} quantum program that is dynamically linked to the system using the program loader client. The mentioned clients and portable experiments are described in Section~\ref{sub:implementation}.

All \ac{artiq} experiments have four execution phases: build, prepare, run, and analyze. The build phase is used to instantiate objects and process arguments. The prepare phase is the first moment where code directly relevant to the experiment can execute. This phase allows experiments to execute code on the host without accessing any devices or data storage. The run phase is the only moment where the experiment has access to devices and data storage. Kernels can only execute in the run phase of the experiment and any data analysis for calibration purposes should also be done here. Finally, the analysis phase is used for the post-experiment analysis of data. The run, prepare, and analyze phases are separated to pipeline experiments and maximize usage of the real-time control system. Our code portability evaluation focuses on the prepare and run phase, which are the two phases directly relevant to the functionality of the experiment. We decided not to add the build and analysis phase to not give ourselves a potentially unfair advantage by including more code in the coverage analysis.

\begin{figure}
    \centering
    \includegraphics[width=\linewidth]{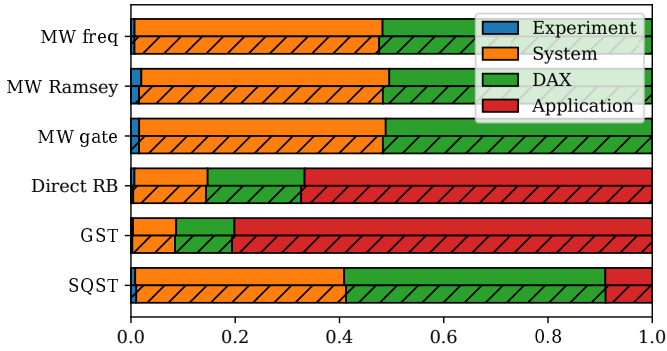}
    \caption{Categorized and normalized proportions of covered statements for the \acs{staq} (solid bars) and \acs{rc} (hatched bars) control software.}
    \label{fig:coverage}
\end{figure}

For our code portability evaluation, we will run the six mentioned experiments on both systems while keeping track of the statement coverage of the prepare and run phase for each experiment. Statement coverage is a technique often used for testing and keeps track of code statements evaluated at least once during program execution.
The resulting data gives insight into the quantity of code that is used during program execution and does not provide information about the execution time spent for each statement.
We measure coverage by simulating our kernel code using the \ac{dax} simulator~\cite{riesebos2022functional} in conjunction with Coverage.py~\cite{coveragepy}.
Our statement coverage data includes statements executed as part of host code, kernels, and \acp{rpc}. For our analysis, we are interested in the coverage of four categories of code: experiment code, system code, \ac{dax} library code, and \emph{application code}. The first two categories are already defined in Section~\ref{sec:architecture} and the third category is self-explanatory. We define application code as high-level and portable code that extends or utilizes the real-time control software to achieve its functionality. For the experiments we chose, application code includes the pyGSTi library and the \ac{sqst} program. Coverage in other supporting libraries, such as \ac{artiq} or the standard library, is not included in this analysis. The coverage results are shown in Figure~\ref{fig:coverage}.

% coverage prep_run staq Experiment [0.007 0.02  0.016 0.007 0.004 0.008] 0.010
% coverage prep_run staq System [0.476 0.476 0.473 0.14  0.083 0.401] 0.341
% coverage prep_run staq AMX [0.517 0.504 0.511 0.186 0.112 0.501] 0.389
% coverage prep_run staq Application [0.    0.    0.    0.667 0.801 0.09 ] 0.260
% coverage prep_run red-chamber Experiment [0.007 0.016 0.016 0.005 0.003 0.01 ] 0.009
% coverage prep_run red-chamber System [0.469 0.468 0.468 0.14  0.082 0.402] 0.338
% coverage prep_run red-chamber AMX [0.524 0.517 0.517 0.183 0.109 0.498] 0.391
% coverage prep_run red-chamber Application [0.    0.    0.    0.673 0.806 0.089] 0.262
% average abs difference in total coverage [ 0.018 -0.011  0.011 -0.01  -0.006  0.006]

The results in Figure~\ref{fig:coverage} show that for all experiments, the proportions of code in each category do not differ much between the \ac{staq} and \ac{rc} system. What can not be seen from the figure is that the total number of statements covered for each experiment does not differ more than 1.8\% between the two systems. Figure~\ref{fig:coverage} shows that the three \ac{mw} calibration experiments have very similar results with 2.0\% or less experiment code, between 46.8\% and 47.6\% system code, and 50.4\% to 52.4\% of \ac{dax} code. The covered statements of \ac{dax} library can mainly be found in its data-processing module, scanning infrastructure, and system initialization-related code.
The Direct \ac{rb} and \ac{gst} experiments both have a large proportion of application code that covers parts of the pyGSTi library. These are procedures used to generate the benchmarking circuits. Both experiments also use pyGSTi for measurement data analysis, but those procedures are not included in our coverage data because they are part of the analysis phase of the experiment. For the remaining portion of covered statements for the Direct \ac{rb} and \ac{gst} experiments, more than half is \ac{dax} library code which includes the code of the client itself and the data processing module.
Finally, the \ac{sqst} program contains 9.0\% and 8.9\% application code, which is the portable \ac{sqst} code itself, for the \ac{staq} and \ac{rc} system, respectively. The \ac{dax} library code mainly includes statements from the program loader client and the data-processing module in addition to the initialization code used by the loader to create and dynamically link the portable program to the system.

The results in Figure~\ref{fig:coverage} show that with a modular software architecture, large portions of covered statements do not have to be system-specific and can be shared as application code or as part of a shared library for system code, such as \ac{dax}. For each unique quantum system, only the experiment code and system code would have to be developed which would significantly reduce the development time. For the code that does need to be developed, most of it is part of the system code which is shared between experiments for a single system and reduces development time even further.

\begin{figure}
    \centering
    \includegraphics[width=\linewidth]{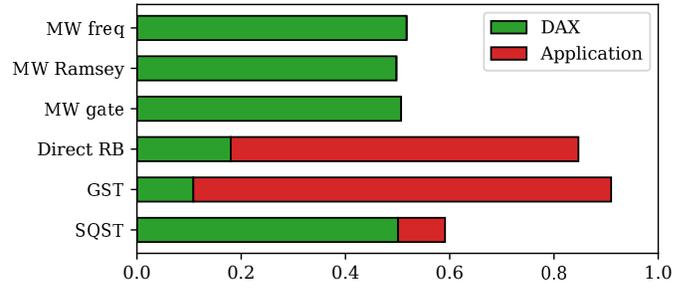}
    \caption{Categorized and normalized proportions of covered statements from \acs{staq} that are shared with \acs{rc}.}
    \label{fig:overlap}
\end{figure}

Only covered statements in the \ac{dax} and application categories in Figure~\ref{fig:coverage} are potentially portable between the \ac{staq} and \ac{rc} systems. We took the coverage data for each experiment and compared how many statements in the \ac{dax} and application categories were covered by both systems. These are the statements that are directly shared between the two systems. The results, which are normalized to the total number of covered statements for \ac{staq}, are shown in Figure~\ref{fig:overlap}.

% overlap prep_run staq [0.517 0.498 0.507 0.847 0.91  0.591] 0.645
% overlap prep_run red-chamber [0.508 0.503 0.501 0.855 0.915 0.587] 0.645
% overlap_2 prep_run staq AMX [0.517 0.498 0.507 0.18  0.108 0.501] 0.385
% overlap_2 prep_run staq Application [0.    0.    0.    0.667 0.801 0.09 ] 0.260

The results in Figure~\ref{fig:overlap} show that even the system-specific \ac{mw} calibration experiments consists of 49.8\% to 51.7\% of shared statements. Data-processing modules and scanning infrastructure add a significant number of covered statements during an experiment and are relatively easy to make portable. By sharing portable modules and services, we achieve portability on the system-code level.
The Direct \ac{rb} and \ac{gst} experiments consist of 84.7\% and 91.0\% of shared statements, respectively. The application code is a major contributor to the proportion of shared statements but portable system code also contributes a significant part. Application code on the quantum operation level is inherently portable and we show that by introducing interfaces and clients, we can successfully connect to application-level software.
Finally, the \ac{sqst} program contains 59.1\% of shared statements of which 9.0\% is application code. The \ac{sqst} program is small and therefore does not contribute a lot of covered statements. The remaining shared statements all originate from the system code.

To further increase the amount of shared code between the \ac{staq} and \ac{rc} systems, we could generalize more modules and services. For example, the \acl{mw} module, the state service, and the \acl{mw} operation service of both systems contain very similar code and could probably be converted to a portable \ac{dax} module or service. So far, we have not done that in favor of flexibility and code simplicity. A module or service part of the system code can easily be modified for testing or mitigating device-related issues. Especially in an academic setting, flexibility can sometimes be more important than code portability. Additionally, portable modules and services are often required to have many configuration and customization capabilities to function correctly for different systems. Hence, portable code is often more complex which might not always be desired. Instead, we keep such modules and services part of the system code and manually "port" them to new systems. Overall development time can still be reduced by porting modules and services while full flexibility is preserved.

\section{Experiments}
\label{sec:experiments}

To demonstrate the capabilities of modular \ac{dax}-based control software and the portability of clients, we used the \ac{rb} client presented in Section~\ref{sub:implementation} to perform benchmarking on two experimental quantum systems.
The \ac{rb} client uses the operation interface which is implemented on both systems by their respective \ac{mw} operation services.
These services utilize a microwave horn to excite a dipole transition between the hyperfine states of \ac{yb171}, which is where the qubit is encoded. This performs $X$ and $Y$ rotations on the Bloch Sphere.  To perform cooling, state preparation, and measurement, the two systems use a 370nm laser to excite the dipole transition between $^2S_{1/2}$ and $^2P_{1/2}$ \cite{olmschenk2007manipulation}.

Before performing coherent operations, very accurate knowledge of the hyperfine frequency difference between the qubit states is needed, along with the Rabi frequency corresponding to oscillations between these states. The hyperfine splitting between the states is very well known \cite{fisk1997accurate}.  However, a strong magnetic field is installed in our systems, slightly altering this value. Assuming we have some knowledge of what the frequency change should approximately be, three calibration experiments are still needed. The first experiment performs a sequence of timed microwave pulses near the qubit transition frequency to get Rabi oscillations.  We then fit this data to get a rough estimate of the Rabi frequency.  The second experiment uses Ramsey interferometry to fine-tune the qubit transition frequency.  This experiment can be done with incrementally smaller frequency ranges to get greater precision of the resonance.  Lastly, we perform increasingly longer sequences of $\pi$-pulse rotations designed to end with the qubit in the ground state in order to fine-tune the Rabi frequency.  

% \subsection{Randomized benchmarking}

After calibration, we perform Direct \ac{rb}, with circuit lengths starting at 1 and scaling up exponentially to 1024. Direct \ac{rb} is a modification of the original \ac{rb} proposal which implements randomized circuits by sampling a system's native gates from a user-provided distribution $\Omega$ \cite{PhysRevLett.123.030503}. For microwave gates on ion trapping systems, the native gate sets are $X$ and $Y$ rotations, and we chose $\Omega$ to be uniform. The sequences are provided by the pyGSTi library \cite{pygsti} using the \ac{dax} \ac{rb} client. For each circuit length, we performed 10 different circuits with 100 samples for each. The circuits were designed such that output was randomized to avoid skewed data because of bias toward a particular outcome.  For example, the detection process in this experimental setup is designed such that the ground state is dark when shining the 370nm laser on the ion. Thus, losing the ion during an experiment would lead to always measuring the ground state.  

\begin{figure}
    \centering
    \includegraphics[width=\linewidth]{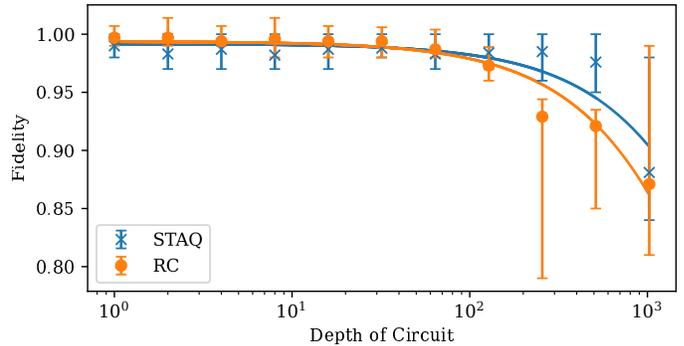}
    \caption{Single-qubit Direct \acl{rb} fidelity results for the \acs{staq} and \acs{rc} system using microwave gates.  Error bars are calculated using the 10th to 90th percentile as boundaries. \acs{rc} starts at higher fidelity due to better \acs{spam} but decays quicker due to lower gate fidelity.}
    \label{fig:RB}
\end{figure}

To demonstrate the flexibility of \ac{dax} and the portability of the \ac{rb} client, we perform Direct \ac{rb} with two different experimental setups: the \ac{staq} and \ac{rc} systems.  Besides the different real-time control systems and devices, the main difference between these two setups is that \ac{staq} is at cryogenic temperatures while \ac{rc} is at room temperature.  However, this shouldn't have any drastic effect on microwave operations and the data between the two systems should be quite comparable.  
% To demonstrate the flexibility of \ac{dax} and the portability of the \ac{rb} client, we perform Direct \ac{rb} with three different implementations of the operation interface in the \ac{staq} system: native microwave gates (service shown in Figure~\ref{fig:staq_system}), microwave gates with first order \ac{sk1} sequences, and microwave gates with first order \ac{bb1} sequences.  \ac{sk1} and \ac{bb1} are compensating pulse sequences designed to mitigate errors in the calibrated Rabi frequency \cite{PhysRevA.70.052318, WIMPERIS1994221}.  The pulse sequence of an arbitrary rotation around the X axis using \ac{sk1} can be written as $R(0, \theta) R(\phi_{\theta}, 2\pi) R(\phi_{\theta}, -2\pi)$, where $\phi_{\theta} = \arccos(-\theta/4\pi)$. The same rotation for \ac{bb1} is defined as $R(0, \theta) R(\phi_{\theta}, \pi) R(3\phi_{\theta}, 2\pi) R(\phi_{\theta}, \pi)$. Rotations around any other axis in the $X/Y$ plane can be performed by adding the required phase offset to each rotation. To emphasize the difference between the operation interfaces, we added an artificial 2\% amplitude error to the calibrated Rabi frequency.  
The results from this experiment can be found in Figure~\ref{fig:RB}. Here, the error per gate $r$ is estimated to be $r = 4(1 - p)/3 = 1.45 \times 10^{-4} \pm 2.58 \times 10^{-5}$ for the \ac{staq} system and $r=2.28 \times 10^{-4} \pm 1.94 \times 10^{-5}$ for the \ac{rc} system, where $p$ is calculated from fitting to the function $P(m) = 0.5 + Bp^m$. This number can also be interpreted as $1 - F_g$, where $F_g$ is the average gate fidelity of the system.
The difference in errors at low circuit depth is simply a result of different \ac{spam} error, while the \ac{staq} system can be seen to overtake \ac{rc} at higher circuit depth due to better gate calibration.  

\acresetall
\section{Conclusion}
\label{sec:conclusion}

We have presented a systematic design strategy and a modular architecture for real-time quantum control software that organizes devices and system-wide functionality in modules and services, respectively. Our architecture supports the development of modular control software and enhances the flexibility and portability of real-time control software.
We implemented a software framework to develop real-time control software based on our proposed architecture, which is part of our open-source library \ac{dax}.
Our evaluation shows that modular control software can reduce the execution time overhead of kernels by 63.3\% on average while not increasing the binary size.
Software portability is achieved on the system level by introducing portable modules, services, and scanning control flow. We achieve application-level portability using interfaces and clients. Our analysis shows that modular control software for two distinctly different systems can share between 49.8\% and 91.0\% of covered code statements.
Finally, we have shown that we can run a portable Direct \ac{rb} experiment on two different ion-trap quantum systems that are fully controlled and calibrated by software based on our framework.

\section*{Acknowledgment}
This work is funded by EPiQC, an NSF Expeditions in Computing (1832377), the Office of the Director of National Intelligence - Intelligence Advanced Research Projects Activity through an ArmyResearch Office contract (W911NF-16-1-0082), the NSF STAQ project (1818914), the U.S. Department of Energy (DOE), Office of Advanced Scientific Computing Research award DE-SC0019294, and DOE Basic Energy Sciences award DE-0019449.

\printbibliography

\end{document}